\documentstyle[12pt,epsf]{article}

\begin{document}        %  DO NOT DELETE OR CHANGE THIS LINE
\pagestyle{empty}
\renewcommand{\thefootnote}{\fnsymbol{footnote}}

%%%%% Substitute your Pub number, month and year in the following:
%%
\begin{flushright}
{\small
SLAC--PUB--8774\\
Feb 2001\\}
\end{flushright}
 
\vspace{.8cm}

%%%%% Title and Author Information:
%%

%%%%% Title and Author Information:
%%
\begin{center}
{\bf\large   
LCD ROOT Simulation and Analysis Tools
\footnote{Work supported 
by Department of Energy contract  DE--AC03--76SF00515 (SLAC).}}

\vspace{1cm}
{\bf
Masako Iwasaki
} \\
\vskip 0.2cm
{\it
Department of Physics, University of Oregon, Eugene,
OR 97403-1274, USA
} \\

\vskip 0.5cm
{\bf
Toshinori Abe
} \\
\vskip 0.2cm
{\it
Stanford Linear Accelerator Center, Stanford CA 94309, USA
} \\

\medskip
\end{center}
 
\vfill

\begin{center}
{\bf\large   
Abstract }
\end{center}
The North American Linear Collider Detector group 
has developed a simulation program package based on the ROOT system.  
The package consists of Fast simulation, the reconstruction of 
the Full simulated data, and physics analysis utilities. 
\vfill

%%%%%%%%%%%%%%%
%% Choose"Presented at," "Contributed to" for conference papers
%% or "Submitted to" for journal papers
%%%%%%%%%%%%%%%
\begin{center} 
{\it Presented at the 5th International Linear Collider Workshop 
(LCWS 2000), 24-28 Oct 2000, Fermilab, Batavia, Illinois, USA} 
\end{center}

\newpage
\pagestyle{plain}

\section{Introduction}
For the various studies of the future Linear Collider experiments, 
the North American Linear Collider Detector group 
(LCD) has constructed a simulation facility. 
It consists of two kinds of detector simulators, Full and Fast simulators, 
designed for detailed detector
studies and physics analyses, respectively.
We use the GISMO package\cite{gismo} for the Full simulation.
Using the digitized outputs from the Full simulation, 
we reconstruct charged tracks and clusters.
The Fast simulation is based on parametrized position and energy smearing, 
and it makes tracks and clusters directly from the generated particle
information.

In the LCD group, we have developed two 
object-oriented program packages based on JAS/Java\cite{JAS} 
and Root/C++\cite{ROOT}, 
for the reconstruction of the Full simulated events, 
the Fast simulation and the event analysis.
In this report, we introduce the simulation and
analysis package based on Root/C++.

\begin{figure}[h] % fig 1
\begin{center}
\epsfysize 2.8in
\epsfbox{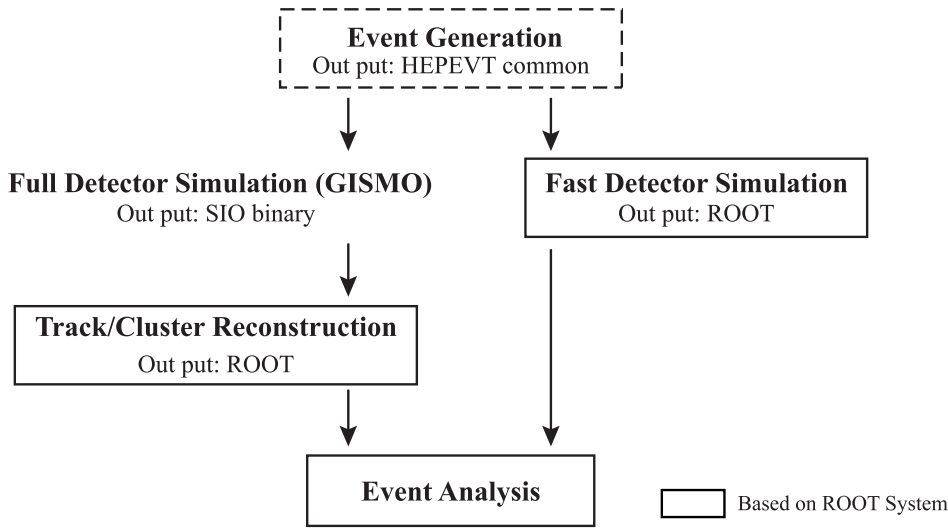}
\end{center}
\caption{The LCD Simulation/Event analysis flow.  
Boxed processes are based on the ROOT system.
Several event generators are also ROOT-based.}
\label{fig1}
\end{figure}

Fig.\ref{fig1} shows the LCD simulation and analysis flow using ROOT.
We use a standard HEPEVT format using the FNAL
StdHep 4.06 I/O package\cite{stdhep}, 
for the generator output and the detector simulator input.
The output format of the Fast simulator is the Root binary, 
it is also possible to be linked with the event analysis directly.
The output format of the Full simulator is the SIO binary and 
is converted to the Root binary. 
Then its reconstruction and the event analysis are done based on the 
LCD Root system.
 
\section{Simulation and Event Reconstruction}
There are two detector models presently under study by the LCD group, a 
Small and a Large detector.
The detector parameters are listed in Table \ref{tab:LandS}.
The Small detector is a compact detector featuring a very strong
magnetic field and solid-state tracking device. 
The calorimeter is also compact
and made with Si-W. The Large detector has a much larger tracker (TPC)
with lower magnetic field and Pb-scintillator calorimeter. 
Both detectors are designed to have a CCD Vertex Detector.
These detector parameters are used for both Full and Fast simulations.
%The simulators are designed to have the flexibility of changing any component
%of the detector parameters.

\begin{table}
\caption{ Detector parameters for LCD Small and Large detectors.}
\label{tab:LandS}
\begin{tabular}{l|c c}
\hline \hline
 & Small & Large \\
\hline
Vertex Detector & CCD & CCD \\
\,\,\,\,\,\,\,\, Impact parameter resolution & 
$3.0 \mu \bigoplus 8.7 \mu / p \sin^{2/3}$ &
$2.5 \mu \bigoplus 8.7 \mu / p \sin^{2/3}$ \\
\hline
Central Tracking & Si micro strips & TPC \\
\,\,\,\,\,\,\,\, Momentum resolution (High)& 
$\delta / P^2 \sim 6 \times 10^{-5}$ &
$\delta / P^2 \sim 5 \times 10^{-5}$ \\
\,\,\,\,\,\,\,\, \hspace{3.4cm} (Low)& 
$\delta P / P \sim 0.4\%$ &
$\delta P / P \sim 0.1\%$ \\
\hline
Electromagnetic Calorimeter & W/Si pads & Pb/scintillator\\
\,\,\,\,\,\,\,\, Barrel Inner Radius &  75 cm & 200 cm \\
\,\,\,\,\,\,\,\, Endcap Inner Z      & 150 cm & 300 cm \\
\,\,\,\,\,\,\,\, Energy resolution &
$\delta E / E \sim 12\%/ \sqrt{E} + 1 \%$ &
$\delta E / E \sim 15\%/ \sqrt{E} + 1 \%$ \\
\,\,\,\,\,\,\,\, Granularity & 20 mrad & 20 mrad \\ 
\hline
Hadron Calorimeter & Cu/scintillator & Pb/scintillator\\
\,\,\,\,\,\,\,\, Barrel Inner Radius & 140 cm & 250 cm \\
\,\,\,\,\,\,\,\, Endcap Inner Z      & 186 cm & 350 cm \\
\,\,\,\,\,\,\,\, Energy resolution &
$\delta E / E \sim 50\%/ \sqrt{E} + 2 \%$ &
$\delta E / E \sim 40\%/ \sqrt{E} + 2 \%$ \\
\,\,\,\,\,\,\,\, Granularity & 60 mrad & 60 mrad \\ 
\hline
Coil Magnet & & \\
\,\,\,\,\,\,\,\, Magnetic field & 6 Tesla & 3 Tesla \\
\,\,\,\,\,\,\,\, Inner Radius & 100 cm (outside EM Cal) & 
376 cm (outside HAD Cal) \\
\hline \hline
\end{tabular}
\end{table}

In the Fast simulation, 
charged particles within the magnet field follow helical trajectories, 
and their momenta and positions are smeared.
Here the charged tracks are expressed by 5 parameters,
and are smeared with a 5$\times$5 covariant error matrix.

Electrons, photons, and hadrons produce clusters in the electromagnetic 
(EM) and hadronic (HAD) calorimeters.
Here, one cluster is made from one particle.
Energies and positions of clusters are smeared.
We assume transverse position resolutions 
of 1cm/$\sqrt{E}$ (electrons and photons) or 5cm/$\sqrt{E}$ (hadrons). 
To consider the detector granularity and cluster width, which is
typically a few units of Moliere radius, 
we merge the clusters when the angular separation between
clusters is less than $\theta_{max}$, where $\theta_{max}$ 
is a size of the detector granularity.

The position of the interaction position is also smeared.
We assumed $\sigma_{x}$=$\sigma_{y}$=2 $\mu$m and 
$\sigma_{z}$=6 $\mu$m. The determination of these values are described
in Ref.\cite{toshi}.

In the reconstruction of the Full simulated data, 
we postpone the track reconstruction. Instead, we make charged tracks
by smearing, using exactly the same procedure as in the Fast simulation, 
but we also apply a minimum tracker-hit cut. 
Calorimeter clusters are made by gathering the hits which are from the 
same particle. Energy and position of the cluster is obtained from 
the energy sum, and the energy-weighted average of associated
Calorimeter hits, respectively.

Comparing the Full and Fast simulations, the most 
significant difference is in the Calorimetry.
It is important to improve the parameterization of 
the Fast simulator Calorimetry to have a
more realistic detector response, which we hope to implement 
in the near future. 

\section{Event Analysis Tools}
For the physics analysis, there are several useful tools
in the LCD Root program. We provide Thrust finding and 3 kinds of Jet finding
(based on JADE, JADE-E and DURHAM algorithms) programs. 

A topological vertex finding algorithm is developed 
in the SLD experiment\cite{ZVTOP}. 
Here the secondary vertices are reconstructed with charged tracks
by searching the space points where track density functions overlap
in the 3D space.
The original SLD prepmort program, called ZVTOP, is  
translated into C++ and several parameters are set for the LCD studies.
Based on the topological vertex finding, we get excellent performances
on heavy-quark (both $b$ and $c$) tagging and charge identification 
of the quark. The details are described in Ref.\cite{toshi}. 
There are physics studies which use the topological vertex
finding in these proceedings\cite{analysis}.

In the energy flow analysis, neutral clusters are selected by 
the absence of a track and cluster association. 
For this analysis, we provide methods which extrapolate 
the particles to the cluster cylindrical radius. 

\begin{figure} [h]% fig 2
\begin{center}
\epsfysize 3.0in
\epsfbox{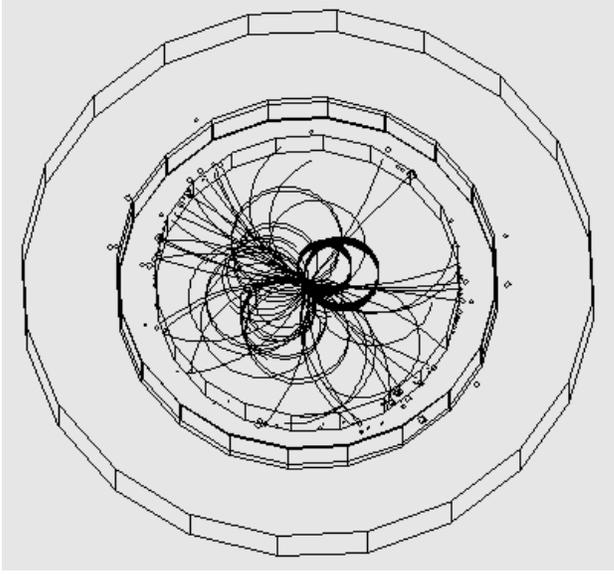}
\end{center}
\caption{ An event display of a simulated 500 GeV $e^+e^-\rightarrow
t\bar{t}$ event. Here the Fast simulator assuming the Large detector
parameters is used.}
\label{fig2}
\end{figure}

As a graphical tool, there is an event display for the LCD Root program. 
An example of the event view for a $e^+e^-\rightarrow t\bar{t}$
event is shown in Fig.\ref{fig2}

\section{Summary}
In this report, we briefly introduce the simulation and analysis tools
based on ROOT for the LCD group. 
Although the tools are constantly being improved, they can be
obtained via the URL,
\begin{tt}
\begin{verbatim}
   http://www-sldnt.slac.stanford.edu/nld/New/Docs/LCD_Root/root.htm.
\end{verbatim}
\end{tt}
Feedback from the users is highly welcomed.

\end{document}